\documentclass[prd,preprintnumbers]{revtex4}

\usepackage{amsmath}
\usepackage{amsfonts}
\usepackage{graphicx}

\begin{document}
\tolerance=5000
\def\pp{{\, \mid \hskip -1.5mm =}}
\def\cL{{\cal L}}
\def\be{\begin{equation}}
\def\ee{\end{equation}}
\def\bea{\begin{eqnarray}}
\def\eea{\end{eqnarray}}
\def\tr{{\rm tr}\, }
\def\nn{\nonumber \\}
\def\e{{\rm e}}
\def\d{{\rm d}}
\preprint{YITP-05-14}

\title{De Sitter cosmology from Gauss-Bonnet dark energy with quantum effects}
% De Sitter cosmology from quantum effects and Gauss-Bonnet dark energy }

\author{Emilio Elizalde}\email{elizalde@ieec.uab.es, elizalde@math.mit.edu}
\affiliation{Consejo Superior de Investigaciones Cient\'{\i}ficas
(ICE/CSIC) \\  Institut d'Estudis Espacials de Catalunya (IEEC) \\
Campus UAB, Facultat de Ci\`encies, Torre C5-Par-2a \\ E-08193 Bellaterra
(Barcelona) Spain}
\author{John Quiroga Hurtado}\email{jquiroga@utp.edu.co}
\affiliation{Department of Physics,\\ Universidad Tecnol\'ogica de
Pereira,\\ Pereira, Colombia}
\author{H\'ector Iv\'an Arcos}\email{hiarcosv@utp.edu.co}
\affiliation{Department of Physics,\\ Universidad Tecnol\'ogica de
Pereira,\\ Pereira, Colombia}
%\date{\today}

\begin{abstract}
A Gauss-Bonnet dark energy model is considered, which is inspired in
string/M-theory and takes also into account quantum contributions.
Those are introduced from a conformal quantum anomaly. The
corresponding solutions for the Hubble rate, $H$, are studied
starting from the Friedmann-Robertson-Walker equation. It is seen
that, as a pure effect of the quantum contributions, a new solution
for $H$ exists in some region, which does not appear in the
classical case. The behavior of all encountered solutions is studied
with care, in particular, the role played by the quantum correction
term---which depends on the number of matter fields---on the
stability of the solutions around its asymptotic value. It is argued
that, contrary to what happens in the classical case, quantum
effects remarkably lead to the realization of a de Sitter stage
which corresponds to the inflation/dark energy stages, even for
positive values of the $f_0$ constant (coupling of the field with
the Gauss-Bonnet invariant).

\end{abstract}


\maketitle

\section{Introduction}

One of the most intriguing questions in today's Physics concerns the
nature of the dark energy known to be present in the Universe we live in. The
existence of such energy---with an almost uniform density distribution
and a substantial negative pressure---which completely dominates all
other forms of matter at present time, is inferred from several astronomical
observations \cite{SuNv}. In particular, according to recent
astrophysical data analysis, this dark energy seems to behave very similarly to a
cosmological constant, which is responsible for the accelerating
expansion of the observable universe. Furthermore, there are strong reasons to
believe that answering this question will have much to do with the
possibility to explain the physics of the very early Universe.

Models of dark energy are abundant. One of the proposed candidates
for it is the so-called phantom field, thus named because it corresponds
to a negative-energy field. The peculiar properties of a phantom scalar (with
negative kinetic energy) in a space with non-zero cosmological
constant have been discussed in an interesting paper by Gibbons
\cite{Gibbons}. As indicated there,  phantom properties bear some
similarity with quantum effects \cite{phtmtr}.
 An important property of the
investigation in \cite{Gibbons} is that it is easily generalizable
to other constant-curvature spaces, such as Anti-de Sitter (AdS)
space. Presently, there is considerable interest in such spaces,
coming in particular from the AdS/CFT correspondence. According to
it, the AdS space may have  cosmological relevance \cite{cvetic},
e.g. by  increasing the number of particles created on a given
subspace \cite{jcap}. It could also be used to study a cosmological
AdS/CFT correspondence \cite{b494}: the study of a phantom field in
AdS space may give us hints about the origin of this field via the
dual description. In the supergravity formulation one may think of
the phantom as of a special renormalization group (RG) flow for
scalars in gauged AdS supergravity. (Actually, such a RG flow may
correspond to an imaginary scalar.)

Another candidate for dark energy is the tachyon \cite{EEQJ1,
snsdo}. This is an unstable field, actually. The interest of models
exhibiting a tachyon is motivated by its role in the
Dirac-Born-Infeld (DBI) action as a description of the D-brane
action \cite{copeland,s1,s2}. In spite of the fact that the tachyon
represents an unstable field, its role in cosmology is generally
still considered to be useful, as a source of dark matter
\cite{snsdo,Gibbons2} and, depending on the form of the associated
potential \cite{Paddy,Bagla,AF,AL,GZ}, it can actually lead to a
period of inflation. On the other hand, it is important to realize
that a tachyon with negative kinetic energy (yet another type of
phantom) can also be introduced \cite{hao}. In that phantom/tachyon
model the thermodynamical parameter $w$ is naturally negative. In
this case, the late-time de Sitter attractor solution is admissible,
and this is one of the main reasons why it can be considered as an
interesting model for dark energy \cite{hao}. Moreover, in order to
understand the role of the tachyon in cosmology it is necessary to
study its effects on other backgrounds, as in the case of an Anti-de
Sitter background \cite{EEQJ1}.

Another remarkable proposal is that the origin of the dark energy
could in fact be related with the problem of the cosmological
constant. One of the most interesting approaches to this paradigm is
modified gravity. Actually, it is not absolutely clear why standard
General Relativity should be trusted at large cosmological scales
(thus, through an enormous range of orders of magnitude). One may
assume that, considering minimal modifications, the gravitational
action contains some additional terms growing slowly in the presence
of decreasing curvature (see, e.g., \cite{capozziello, NOPRD, ln,
tr, sasaki,sami,string} and, for a review, see \cite{GB2}), which
could be responsible for the current acceleration. Within such
scenario one of the most accepted approaches is the model of
modified Gauss-Bonnet (GB) gravity. In this model additional terms
in the gravitational action are introduced by means of a function
which depends on the scalar curvature, $R$, and on the Gauss-Bonnet
invariant, $G$ \cite{GB1}. It is possible to demonstrate that such
models lead to a plausible effective cosmological constant,
quintessence, or phantom eras. From these results  one can conclude
that, concerning its role as a gravitational alternative for DE,
modified GB gravity may become a very strong candidate
\cite{GB-gravity}.

In the present paper we will consider a Gauss-Bonnet model for
gravity, together with the first contributions coming from quantum
effects. A stringy correction will be added to the ordinary General
Relativity action, that is, a term proportional to the GB invariant
$G$. We will duely take into account the quantum effects coming from
matter, and their influence on the stability of the de Sitter
universe will be carefully discussed.

\section{THE MODEL}
We  consider the model consisting of a scalar field, $\phi$, coupled
with gravity in a rather non-trivial way and, as advanced, we will
also take into account quantum effects.

As a stringy correction, the term proportional to the Gauss-Bonnet
invariant, $G$, is written as:
 \be \label{GB1} G=R^2 - 4 R_{\mu\nu} R^{\mu\nu} +
R_{\mu\nu\rho\sigma} R^{\mu\nu\rho\sigma}\ . \ee The starting
action is given by \be \label{GB2} S=\int d^4x \sqrt{-g}\left\{
\frac{1}{2\kappa^2}R
    - \frac{\gamma}{2}\partial_\mu \phi \partial^\mu \phi
    - V(\phi) + f(\phi) G\right\}\ .
\ee Here $\gamma=\pm 1$. Action (\ref{GB2}) has been proposed as a
stringy dark energy model in \cite{sasaki}. It is interesting to
note that one can also add to this system $R^3$ and $R^4$ stringy
corrections \cite{sami, sami1} (for a general introduction to
modified gravity, see \cite{GB2, snsd1}). Moreover, action
(\ref{GB2}) is able to solve the initial singularity problem
\cite{ART, nick} For the canonical scalar, $\gamma=1$ and, at least
when the GB term is not included, the scalar behaves as a phantom
only when $\gamma=-1$ \cite{caldwell}, showing in this case
properties similar to those of a quantum field \cite{phtmtr}. In
analogy with the model in \cite{coupled}---where also a non-trivial
coupling of the scalar Lagrangian with some power of the curvature
was considered---one may expect that a GB coupling term of this kind
can be relevant for the explanation of the dark energy dominance
nowadays.

Doing the same as in Ref.~\cite{sasaki}, by varying over $\phi$, one
obtains \be \label{GB3} 0=\gamma \nabla^2 \phi - V'(\phi) + f'(\phi)
G\ . \ee On the other hand, performing the variation over the metric
$g_{\mu\nu}$ (as in \cite{gcsnsdosz}), gives \bea \label{GB4} 0&=&
\frac{1}{\kappa^2}\left(- R^{\mu\nu} + \frac{1}{2} g^{\mu\nu}
R\right)
      + \gamma \left(\frac{1}{2}\partial^\mu \phi \partial^\nu \phi
      - \frac{1}{4}g^{\mu\nu} \partial_\rho \phi \partial^\rho \phi \right)
   + \frac{1}{2}g^{\mu\nu}\left( - V(\phi) + f(\phi) G \right) \nn
&&    -2 f(\phi) R R^{\mu\nu} + 2 \nabla^\mu \nabla^\nu
\left(f(\phi)R\right)
      - 2 g^{\mu\nu}\nabla^2\left(f(\phi)R\right) \nn
&& + 8f(\phi)R^\mu_{\ \rho} R^{\nu\rho}
      - 4 \nabla_\rho \nabla^\mu \left(f(\phi)R^{\nu\rho}\right)
      - 4 \nabla_\rho \nabla^\nu \left(f(\phi)R^{\mu\rho}\right) \nn
&& + 4 \nabla^2 \left( f(\phi) R^{\mu\nu}  \right) + 4g^{\mu\nu}
\nabla_{\rho} \nabla_\sigma \left(f(\phi) R^{\rho\sigma} \right)
   - 2 f(\phi) R^{\mu\rho\sigma\tau}R^\nu_{\ \rho\sigma\tau}
+ 4 \nabla_\rho \nabla_\sigma \left(f(\phi)
R^{\mu\rho\sigma\nu}\right) \, . \eea And using the identities
obtained from the Bianchi identity, \bea \label{GB5} \nabla^\rho
R_{\rho\tau\mu\nu}&=& \nabla_\mu R_{\nu\tau} - \nabla_\nu
R_{\mu\tau}\ ,\nn \nabla^\rho R_{\rho\mu} &=& \frac{1}{2} \nabla_\mu
R\ , \nn \nabla_\rho \nabla_\sigma R^{\mu\rho\nu\sigma} &=& \nabla^2
R^{\mu\nu} - {1 \over 2}\nabla^\mu \nabla^\nu R +
R^{\mu\rho\nu\sigma} R_{\rho\sigma}
      - R^\mu_{\ \rho} R^{\nu\rho} \ ,\nn
\nabla_\rho \nabla^\mu R^{\rho\nu} + \nabla_\rho \nabla^\nu
R^{\rho\mu} &=& {1 \over 2} \left(\nabla^\mu \nabla^\nu R +
\nabla^\nu \nabla^\mu R\right)
      - 2 R^{\mu\rho\nu\sigma} R_{\rho\sigma}
+ 2 R^\mu_{\ \rho} R^{\nu\rho} \ ,\nn \nabla_\rho \nabla_\sigma
R^{\rho\sigma} &=& {1 \over 2} \Box R \ , \eea one can rewrite
(\ref{GB4}) as \bea \label{GB4b} 0&=& \frac{1}{\kappa^2}\left(-
R^{\mu\nu} + \frac{1}{2} g^{\mu\nu} R\right)
      + \gamma \left(\frac{1}{2}\partial^\mu \phi \partial^\nu \phi
      - \frac{1}{4}g^{\mu\nu} \partial_\rho \phi \partial^\rho \phi \right)
   + \frac{1}{2}g^{\mu\nu}\left( - V(\phi) + f(\phi) G \right) \nn
&&    -2 f(\phi) R R^{\mu\nu} + 4f(\phi)R^\mu_{\ \rho} R^{\nu\rho}
   -2 f(\phi) R^{\mu\rho\sigma\tau}R^\nu_{\ \rho\sigma\tau}
+4 f(\phi) R^{\mu\rho\sigma\nu}R_{\rho\sigma} \nn && + 2 \left(
\nabla^\mu \nabla^\nu f(\phi)\right)R
      - 2 g^{\mu\nu} \left( \nabla^2f(\phi)\right)R
   - 4 \left( \nabla_\rho \nabla^\mu f(\phi)\right)R^{\nu\rho}
      - 4 \left( \nabla_\rho \nabla^\nu f(\phi)\right)R^{\mu\rho} \nn
&& + 4 \left( \nabla^2 f(\phi) \right)R^{\mu\nu} + 4g^{\mu\nu}
\left( \nabla_{\rho} \nabla_\sigma f(\phi) \right) R^{\rho\sigma} -
4 \left(\nabla_\rho \nabla_\sigma f(\phi) \right)
R^{\mu\rho\nu\sigma} \ . \eea The above expression is valid in
arbitrary spacetime dimensions. In four dimensions, the terms
proportional to $f(\phi)$ without derivatives are canceled among
themselves, and vanish, since the GB invariant is a total derivative
in four dimensions.

Quantum effects can be included by taking into account the
contribution of  the conformal anomaly, $T_A$, defined as follows:
\be \label{OVII} T_A=b\left(F+{2 \over 3}\Box R\right) + b' G +
b''\Box R\ , \ee where $F$ is the square of a 4d Weyl tensor and $G$
is the Gauss-Bonnet curvature invariant. They are given by \bea
\label{GF} F&=& (1/3)R^2 -2 R_{ij}R^{ij}+ R_{ijkl}R^{ijkl}\,, \nn
G&=&R^2 -4 R_{ij}R^{ij}+ R_{ijkl}R^{ijkl}\,. \eea In general, with
$N$ scalar, $N_{1/2}$ spinor, $N_1$ vector fields, $N_2$ ($=0$ or
$1$) gravitons, and $N_{\rm HD}$ higher-derivative conformal
scalars, the coefficients $b$ and $b'$ take the form \bea \label{bs}
\hspace*{-2.2em} && b={N +6N_{1/2}+12N_1 + 611 N_2 - 8N_{\rm HD}
\over 120(4\pi)^2}\,, \nn \hspace*{-2.2em} && b'=-{N+11N_{1/2}+62N_1
+ 1411 N_2 -28 N_{\rm HD} \over 360(4\pi)^2}\,. \eea We have that
$b>0$ and $b'<0$ for usual matter, except for higher derivative
conformal scalars. Notice that $b''$ can be shifted by a finite
renormalization of the local counterterm $R^2$, thus  $b''$ can be
arbitrary.

In terms of the corresponding energy density \cite{nojiri,
tsujikawa}, $\rho_A$, and of the pressure density, $p_A$, $T_A$ is
given by $T_A=-\rho_A + 3p_A$. Using the energy conservation law in
the Friedmann-Robertson-Walker (FRW) universe, \be \label{CB1}
\dot{\rho}_A+3 H\left(\rho_A + p_A\right)=0\,, \ee we can eliminate
$p_A$, as \be \label{CB2} T_A=-4\rho_A -\dot{\rho}_A/H \,. \ee This
yields the following expression for $\rho_A$: \bea \label{CB3}
\hspace*{-0.4em} \rho_A&=& -\frac{1}{a^4} \int \d t\, a^4 H T_A \nn
\hspace*{-0.4em}&=& -\frac{1}{a^4} \int \d t\, a^4 H \Bigl[-12b
\dot{H}^2 + 24b' (-\dot{H}^2 + H^2 \dot{H} + H^4) \nn
\hspace*{-0.5em}& &- (4b + 6b'')\left(\dddot{H} + 7 H \ddot{H} +
4\dot{H}^2 + 12 H^2 \dot{H} \right) \Bigr]\,. \eea

Now, for the FRW universe metric,
 \be
\label{GB5b} ds^2=-dt^2 + a(t)^2 \sum_{i=1}^3 \left(dx^i\right)^2\
, \ee
taking into account the contribution from the quantum anomaly, $T_A$,
the equation of motion (\ref{GB4}) becomes
\be \label{GB7} 0=-\frac{3}{\kappa^2}H^2 + \frac{\gamma}{2}{\dot
\phi}^2 + V(\phi) - 24 \dot \phi f'(\phi) H^3 + \rho_A\ . \ee
An equation of this sort (\ref{GB7}) was also obtained  for dilatonic
gravity \cite{brevik99,quiroga7}.
On the other hand, Eq.~(\ref{GB2}) becomes \be
\label{GB8} 0=-\gamma\left(\ddot\phi + 3H\dot\phi\right) -
V'(\phi) + 24 f'(\phi) \left(\dot H H^2 + H^4\right)\ . \ee
In the above equations it has been assumed that $\phi$ only depends on
time.

Postulating the de Sitter solution, i.e. looking for solutions of
the form $a=a_0\exp{(Ht)}$, where $H$ and the scalars are constant,
we have for $\rho_A$ that \be \label{CB3-1} \rho_A = -6b'H \,.\ee
After this, Eq.~(\ref{GB7}) becomes \be \label{GB9}
0=-\frac{3}{\kappa^2}H^2 + V(\phi) - 6b'H\ . \ee And, since we are
looking for solutions with $H=$ const, we  are now interested in
solving the equations of motion in the form \bea \label{GB10}
-V'(\phi)+24f'(\phi)(\dot{H}H^2+H^4)&=&0\,,\\
\label{GB11} -\frac{3}{\kappa^2}H^2+V(\phi)-6b'H&=&0\,.\eea

We  may now consider the case when $V(\phi)$ and $f(\phi)$ are given
as exponents, with some constant parameters $V_0\,, f_0
\,\mbox{and}\, \phi_0$, \be V=V_0\ \e^{-\frac{2\phi}{\phi_0}}\
,\quad f(\phi)=f_0 \ \e^{\frac{2\phi}{\phi_0}}\, , \ee these are
string-inspired values for the potentials. Assuming the scale factor
to behave as $a=a_0\exp{(Ht)}$, we get the following equations of
motion \bea \label{GB12}
\frac{2}{\phi_0}V_0{\e^{-\frac{2\phi}{\phi_0}}}+ 48\left(\frac{f_0}{\phi_0}\right) \e^{\frac{2\phi}{\phi_0}}H^4&=&0 \,,\\
\label{GB13}
-\frac{3}{\kappa^2}H^2+V_0\e^{-\frac{2\phi}{\phi_0}}-6b'H&=&0\,.\eea

Now, combining Eqs.~(\ref{GB12}) and (\ref{GB13}) we find the
following equation for $H$: \be \label{GB14}H^3+\left(
\frac{\e^{-\frac{2\phi}{\phi_0}}}{8f_0\kappa^2}\right)H+\frac{b'}{4f_0}\e^{-\frac{2\phi}{\phi_0}}=0\,.\ee
If we switch off the quantum effects ($b'=0$), then we find for
$H^2$: \be\label{GB20} H^2=-\frac{V}{8V_0f_0\kappa^2}\,.\ee This is
the pure classical solution, which was obtained in \cite{sasaki}.
This solution may require the coupling $f_0$ to be negative,
otherwise the solution may not exist. Furthermore, since the Hubble
rate can be determined by the initial condition ($H=H_0$), the
choice of $\phi_0$ is fully arbitrary.

Returning to Eq.~(\ref{GB14}), in looking for a solution which takes
into account quantum effects, we easily find a numerical one by
rewriting Eq.~(\ref{GB14}) in terms of the potential $V$, as
follows, \be\label{GB21}
H^3+\frac{V}{8f_0V_0\kappa^2}H+\frac{b'V}{4V_0f_0}=0\,.\ee This
equation has the form of a reduced cubic equation and thus it is
possible to find a solution by using Cartan's formula. The
discriminant of the cubic equation is  \be\label{GB15} D=
\frac{V^3}{(24)^3f_0^3V_0^3\kappa^6}\left(
1+\frac{216b'^2f_0V_0\kappa^6}{V}\right)\,.\ee As seen from
(\ref{GB15}), for the cubic equation (\ref{GB14}) the sign of the
discriminant depends on the sign of $f_0$, since we define $V_0$ to
be positive.

For the analysis of the solutions of Eq.~(\ref{GB21}) in this case,
we will better consider numerical solutions and look for the
behavior of  the Hubble rate $H$ as a function of the potential $V$.
Let us consider, as a first case, the solution with the following
values for the parameters: $f_0=-0.5,\ V_0=1,\ b'=-0.5$, and
$\kappa=16\pi G$, where $G=6.63\times 10^{-8}$ is the gravitational
constant. For those, the behavior of $H$ is illustrated in Fig.~1.
\begin{figure}[ht]%\vskip2cm
\begin{center}
\includegraphics[height=8cm, width= 8cm]{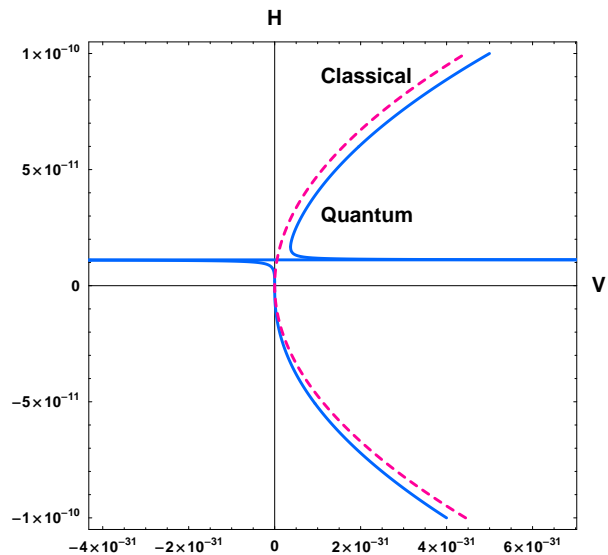}
\caption{Behavior of the Hubble rate $H$ as a function of the
potential $V$, for several values of $V$.}
\label{fig1}\end{center}\end{figure} We immediately see from this
figure that we indeed have three possible ways to define $H$ as a
function of the potential $V$. That is, of course, a consequence of
the three possible solutions of the cubic equation (\ref{GB21}). We
will be interested only in the case when $H$ has positive values for
positive values of the potential $V$.

As seen in the plot, the Hubble rate $H$ is growing as the potential
grows, starting from some minimal positive value. From another side,
it is important to remark that there is a positive solution for $H$
which is very close to the classical solution (dashed line), up to
the point where the continuous-line solution tends asymptotically to
a very small value. This case clearly shows that quantum effects are
only a perturbation to the classical solution. It is also clear from
Fig.~\ref{fig1} the interesting feature that the asymptotic value of
$H$ depends on the quantum correction term $b^\prime$, which on its
turn depends on the number of fields $N$. This means that the
asymptotic value attains a greater positive value if $N$ is larger.
In the absence of  the $b^\prime$ term this discontinuous behavior
of the solution for $H$ disappears. This leads to the conclusion
that the instability of the solution around its asymptotic value is
exclusively due to quantum effects.

It is remarkable that there exists, owing to the quantum effects,
another possible solution for $H$ with a positive value for $f_0$.
This solution exists in a very small region, as a pure effect of
quantum contributions since it turns out that in the classical case
\cite{sasaki} it is not possible to obtain any positive solutions
for $H$---as seen from Eq.~(\ref{GB20}). In Fig.~2 the situation we
have just discussed is shown, corresponding to a positive value for
$f_0$.
\begin{figure}[h]%\vskip2cm
\begin{center}
\includegraphics[height=8cm, width= 8cm]{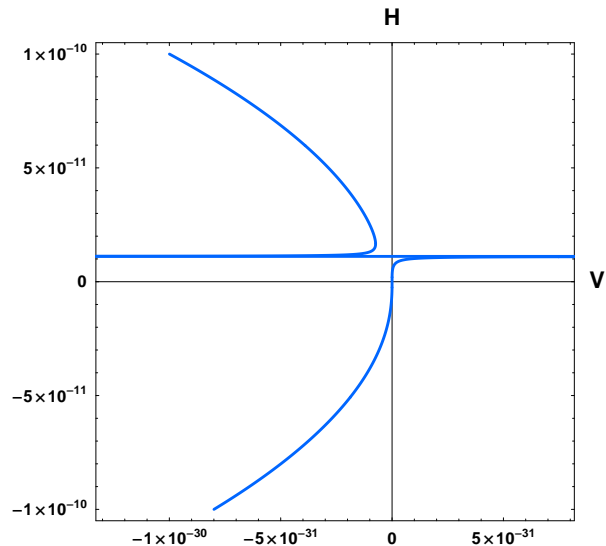}
\caption{Behavior of the Hubble rate $H$ for several values of $V$, with
a positive coupling
of the scalar field with the Gauss-Bonnet invariant: $f_0>0$.}
\label{fig2}\end{center}\end{figure}% \vskip2cm

\section{CONCLUSIONS}

To summarize, a number of quite interesting conclusions can be drawn
from the present study of the influence of a combination of quantum
effects and modified gravity as a possible way for interpreting dark
energy in an accelerated inflationary de Sitter universe. In
particular, it is expected that for large values of the curvature
($R=6\dot{H}+12H^2$) the de Sitter epoch thus described corresponds
to early-time inflation.
\begin{figure}[h]%\vskip2cm
\begin{center}
\includegraphics[height=8cm, width= 8cm]{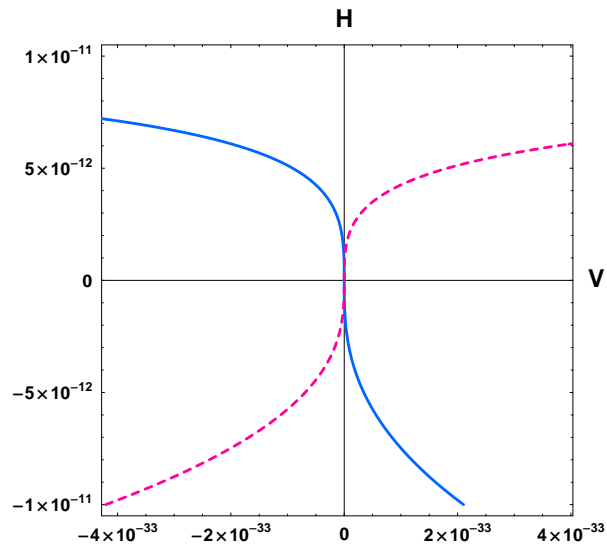}
\caption{Behavior of the Hubble rate $H$ near its value at the dark
energy stage. The dashed line corresponds to the positive coupling case,
$f_0>0$, and the continuous line to the negative case, $f_0<0$.}
\label{fig3}\end{center}\end{figure}% \vskip2cm

From another point of view, for small values of curvature [$R\sim
(10^{-33}$eV$)^2$] we expect that the de Sitter era realized in this
model will correspond to the dark energy stage. However, as seen in
Fig.~\ref{fig3}, this occurs for both positive and negative values
of the potential, depending on the sign of $f_0$. This result
clearly shows that, contrary to what happens in the classical case,
quantum effects lead in fact to the realization of the de Sitter era
corresponding to the dark energy stage, even for positive values of
the $f_0$ constant. Such situation may potentially have interesting
cosmological consequences.

In a similar fashion, the coupling of GB dark energy with $F(R)$
gravity could be considered, too. This should not represent a big
problem, in principle, owing to the fact that the curvature is
constant on the solutions. That possibility deserves further
investigation.

\medskip

\noindent {\bf Acknowledgments}

We are grateful to S.D. Odintsov for very helpful discussions. The
research of JQH and HIA has been supported by Grants-in-Aid for
Scientific Research, No 3-07-05 at the Universidad Tecnol\'ogica de
Pereira, Colombia. The research of EE was partly done while on leave
at the Department of Physics \& Astronomy, Dartmouth College, 6127
Wilder Laboratory, Hanover, NH 03755, USA. EE was supported in part
by MEC (Spain), project PR2006-02842, and by AGAUR (Gene\-ra\-litat
de Ca\-ta\-lu\-nya), grant 2007BE-1003 and contract 2005SGR-00790.

\end{document}